\documentclass[%
 reprint,
 amsmath,amssymb,
 aps,
 onecolumn,
]{revtex4-2}

\usepackage{graphicx}
\usepackage{dcolumn}
\usepackage{bm}
\usepackage[mathlines]{lineno}
\usepackage{txfonts}

\begin{document}

\preprint{APS/123-QED}

\title{\textbf{ Prompt acceleration of  the $\mu^+$ beam in a donut wakefield driven by a shaped Laguerre-Gaussian laser pulse  }}

\author{Xiao-Nan Wang}
 \altaffiliation[Also at ]{Physics and Space Science College, China West Normal University}
 \author{Xiao-Fei Lan}%
 \email{lan-x-f@163.com}
\affiliation{%
Physics and Space Science College, China West Normal University, Nanchong 637009, China
}%

\author{Yong-Sheng Huang}
\email{huangys82@ihep.ac.cn}
\affiliation{
 Institute of High Energy Physics, CAS, Beijing 100049, China
}%

\author{You-Ge Jiang}
 \altaffiliation[Also at ]{Physics and Space Science College, China West Normal University}

\author{Hao Zhang}
\affiliation{%
 Department of Physics, National University of Defense Technology, Changsha 410073, China
}%

\begin{abstract}
The recent experimental data of anomalous magnetic moments strongly indicate the existence of new physics beyond the standard model. An energetic $\mu^+$ beam is a potential option to the expected neutrino factories, the future muon colliders and the $\mu$SR(the spin rotation, resonance and relaxation) technology. It is proposed a prompt acceleration scheme of the $\mu^+$ beam in a donut wakefield driven by a shaped Laguerre-Gaussian (LG) laser pulse. The forward part of the donut wakefield can accelerate and also focus positive particle beams effectively. The LG laser is shaped by a near-critical-density plasma. The shaped LG laser has the shorter rise time and can enlarge the acceleration field. The acceleration field driven by a shaped LG laser pulse is six times higher than that driven by a normal LG laser pulse. The simulation results show that the $ \mu^+$ bunch can be accelerated from $200\mathrm{MeV}$ to 2GeV and the transversal size of the $\mu^+$ bunch is also focused from initial $\omega_0=5\mu m$ to $\omega=1\mu m$ within several picoseconds.
\end{abstract}

\maketitle

Recently, there are a booming interest on the exploration of new physics beyond the Standard Model by the $\mu^+(\mu^-)$ rare decay \textsuperscript{\cite{berger2014mu3e1}}\textsuperscript{\cite{bartoszek2015mu2e2}}\textsuperscript{\cite{kutschke2009mu2e3}}\textsuperscript{\cite{grassi2005meg4}}\textsuperscript{\cite{kuno2013search5}} and the anomalous magnetic moment \textsuperscript{\cite{farley20044756}}\textsuperscript{\cite{charpak1961measurement7}}\textsuperscript{\cite{bailey1979final8}}. As a unstable particle, $\mu^+(\mu^-)$ with the rest mass $m_\mu =207m_e$ and the relatively long rest lifetime $\tau= 2.2\mu s$ has widely applications. An energetic $\mu^+(\mu^-)$ beam  is a competitive candidate for the expected neutrino factories\textsuperscript{\cite{neutrinofac9}}\textsuperscript{\cite{cao2014muon10}} and the future muon colliders\textsuperscript{\cite{ankenbrandt1999status11}}\textsuperscript{\cite{barger1995s}}. For the $\mu$SR(the spin rotation, resonance and relaxation) technology \textsuperscript{\cite{wang2001nuclear12}}\textsuperscript{\cite{sonier2000musr13}}\textsuperscript{\cite{amato1997heavy14}}\textsuperscript{\cite{roduner2012positive15}} , an energetic $\mu^+$ beam can through the wall of the container to research the materials in complex environments. However, until now, there are two main types of $\mu^+(\mu^-)$ sources: the high-energy low-flux cosmic muon source \textsuperscript{\cite{olbert1954production20}}\textsuperscript{\cite{bose1944cosmic21}} and the low-energy muon sources produced by traditional accelerator\textsuperscript{\cite{miyake2009j22}}\textsuperscript{\cite{morenzoni2000low23}}\textsuperscript{\cite{matsuzaki2001riken24}}. The low-energy muon has short lifetime and the cosmic muon source has so small flux to be used for the study of new physics.

  The plasma-based accelerators\textsuperscript{\cite{leemans2014multi25}}\textsuperscript{\cite{chen1985acceleration26}}\textsuperscript{\cite{litos201427}}\textsuperscript{\cite{hogan2010plasma28}} offer the extremely high acceleration fileds of several hundred GV/m with potential applications for high-energy physics and particle sources. Recently, the $\mu^-$ beam can be accelerated to GeV in a laser plasma wakefield\textsuperscript{\cite{zhang2018all29}}. However, the $\mu^+$ beam acceleration in the plasma wakefield is unexplored. The $\mu^+$ beam will be defocused by the transversal wakefield driven by a general gauss laser pulse. We propose a prompt acceleration scheme for the $\mu^+$  beam in a donut wakefield\textsuperscript{\cite{vieira2014nonlinear30}}\textsuperscript{\cite{mendoncca2014donut31}}\textsuperscript{\cite{zhang2016acceleration32}} driven by a shaped LG laser pulses\textsuperscript{\cite{allen1992orbital33}}. The forward part of the donut wakefield can focus and accelerate the $\mu^+$ beam, simultaneously.

In this paper, we use the three-dimensional particle-in-cell (PIC) simulation of Epoch 3D\textsuperscript{\cite{arber2015contemporary34}} and a sample model to demonstrate that the $\mu^+$ bunch can be focused and accelerated in the donut wakefield driven by a LG laser. The simulation results show that the transversal size of the $\mu^+$ beam is focused from initial $\omega_0=5\mu m$ to $\omega=1\mu m$ within several picoseconds. But the energy gain of a $300\mathrm{MeV}$ $\mu^+$ bunch is about $200\mathrm{MeV}$. The LG laser can be shaped by a near-critical-density plasma. The shaped LG laser has the shorter rise time and can push the plasma electrons form an electron sheath with larger $\sigma_e \equiv n_{sh}\Delta x_{n_{sh}}$ at the front of the donut wakefield, where $n_{sh}$ and $\Delta x_{n_{sh}}$ are the density and the width of the electron sheath. The acceleration field $E_x$ is proportional to the $\sigma_e$. The simulation results show that in the donut wakefield driven by the shaped LG laser, the $300\mathrm{MeV}$ $\mu^+$ bunch's energy gain is about $1.5\mathrm{GeV}$ and the transversal size of the $\mu^+$ bunch is also focused from initial $\omega_0=5\mu m$ to $\omega=1\mu m$ within several picoseconds.

\begin{figure*}
\includegraphics[scale=0.6]{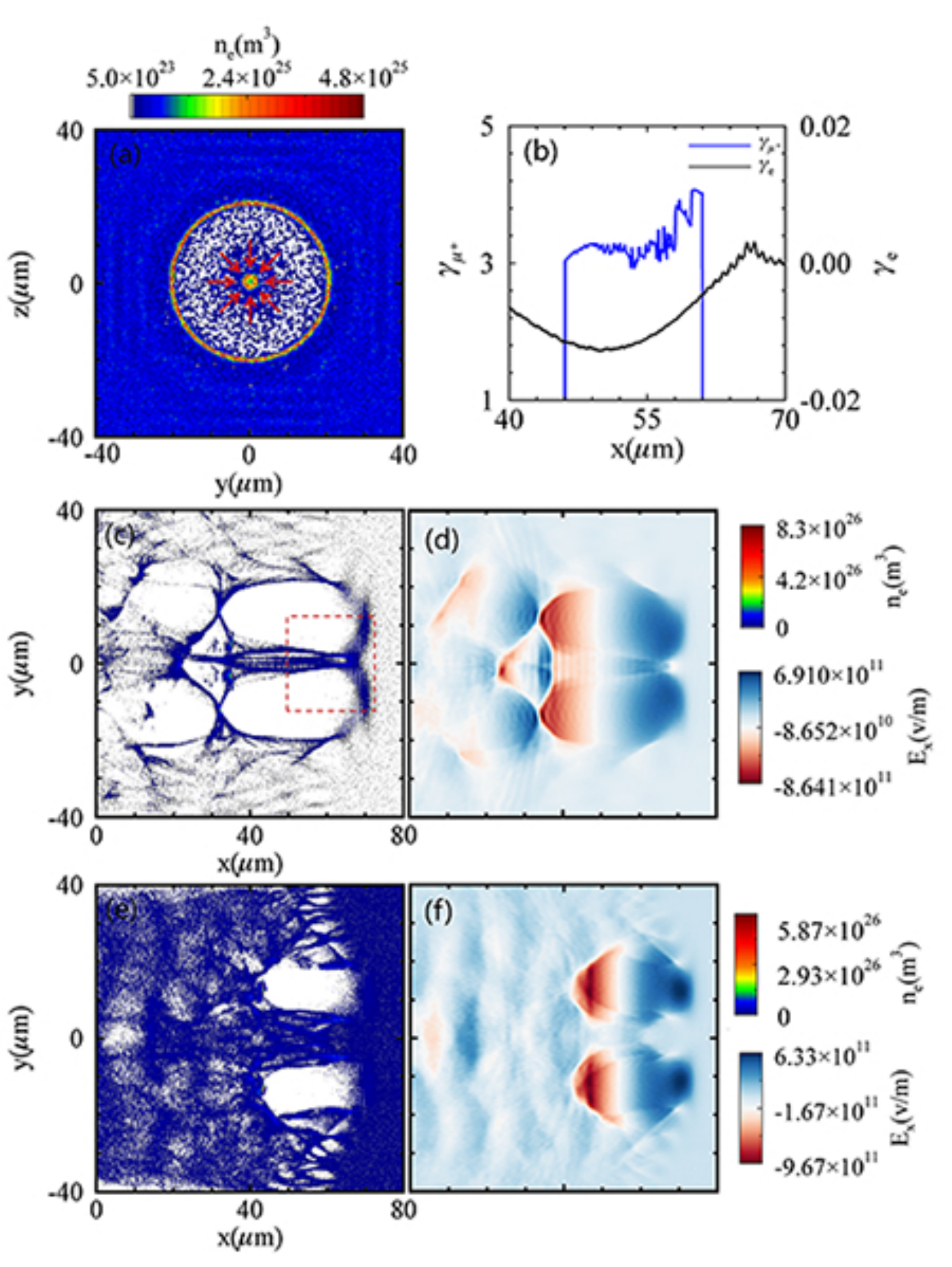}
\caption{\label{fig1} \small \textbf{ Simulation results illustrating the focusing (a), (b) and accelerating (c), (d) fields for the $\mu^+$ beam in a nonlinear donut wakefield.} The laser and plasma parameters of (a-d) are same with the those of type A shown in Method. (a) is the transversal slice of the donut wakefield at the plane of $x=60\mu m$. The red arrow in (a) is the direction of the transversal electric field produced by the electron column inner the donut wakefield.  (b) shows the lineouts of the electron's and $\mu^+$'s relativistic factors at the center line of the donut wakefield. (c) is the longitudinal slice of the electron density at $z=0\mu m$. (d) is the distribution of the longitudinal wakefield $E_x$ corresponding to (c). (e) shows the longitudinal slice of the donut wakefield at $z=0\mu m$ in a plasma with higher density $n_e=1\times 10^{25}m^3$. This donut wakefield is closes to the onset of the nonlinear regime. (f) is the distribution of the $E_x$ corresponding to (e).}
\end{figure*}

The transversal distribution of the LG laser pulse's intensity can be expressed as:
\begin{equation}
a_{l,p}(r,\theta)=a_0 (\frac{c_{l,p}}{\omega})  (\frac{\sqrt{2}r}{\omega})^l  exp(\frac{-r^2}{\omega^2}) exp(-il\theta)   L_p^l(\frac{2r^2}{\omega ^2}), \label{eq1}
\end{equation}
where $a_0$ is the maximum normal intensity, $c_{l,p}$ is the  normalizing factor, $\omega$ is the laser spot size, $L_p^l (\xi)$ is the generalized Laguerre polynomial. In our simulations, the Gaussian mode of the laser beam is the $LG_{1,0}$ model, which is expressed as:
\begin{equation}
a_{1,0} (r,\theta)=a_0 \frac{c_{1,0}}{\omega} \frac{\sqrt{2}  r}{\omega}  exp(\frac{-r^2}{\omega^2})  exp(-il\theta).   \label{eq2}
\end{equation}
  Eq. (\ref{eq2}) shows that the laser intensity is equal to zero at $r=0$. Due to the pondermotive force of the LG laser pulse, the plasma electrons will be squeezed into the center axis and form an electron column. This electron column has the higher density than the background electrons . The pondermotive force of the LG laser pulse also exclude the plasma electrons around. The plasma protons are stable. The uniform protons pull the excluded electrons back to the center axis. The donut wakefield is formed.

 The $\mu^+$ beam can be accelerated and focused at the center line of the donut wakefield in a nonlinear scheme\textsuperscript{\cite{lu2006nonlinear35}}\textsuperscript{\cite{lu2007generating36}} driven by the LG laser. Figure \ref{fig1}(a) shows the transversal density slice of the wakefield driven by the LG laser pulse, which looks like a donut. The transversal electric field of the electron column inner the donut can prevent the defocusing of the $\mu^+$ beam. Figure \ref{fig1}(b) shows that the relativistic factor of the $\mu^+$ is larger than that of the electron column. The magnetic field of the electron column inner the donut also can provide the focusing force for the $\mu^+$ beam.  Figure \ref{fig1}(c) shows the longitudinal density slice of the donut wakefield, which is looks like that there are two spherical wakefields. Due to the radian of the electron sheath at the front of the two spherical wakefields, the electric filed of the electron sheath has a component in the x direction, which also can provide the focusing force for the $\mu^+$ beam. The electron column still provides the main focusing force for the $\mu^+$ beam. The donut wakefield in Figure \ref{fig1}(c), (e) is driven by the same LG laser pulse in the plasmas with the different density $n_e=5\times10^{24}m^3$ and $n_e=1\times10^{25}m^3$, respectively. Figure \ref{fig1}(c) shows that in the nonlinear scheme, there is an overlapped range of the two spherical wakefields. Figure \ref{fig1}(d) shows the longitudinal wakefield $E_x$ hardly varies with r, namely $\partial E_x/\partial r\simeq0$. Figure \ref{fig1}(e) shows that in the onset of nonlinear scheme, there is no overlapped range of the two spherical wakefields. The longitudinal wakefield is close to zero at $r=0$ shown in Figure \ref{fig1}(f). Therefor, only the nonlinear donut wakefield can accelerate and focus the $\mu^+$ beam at the center line of the donut wakefield.
  \begin{figure*}
\includegraphics[scale=0.5]{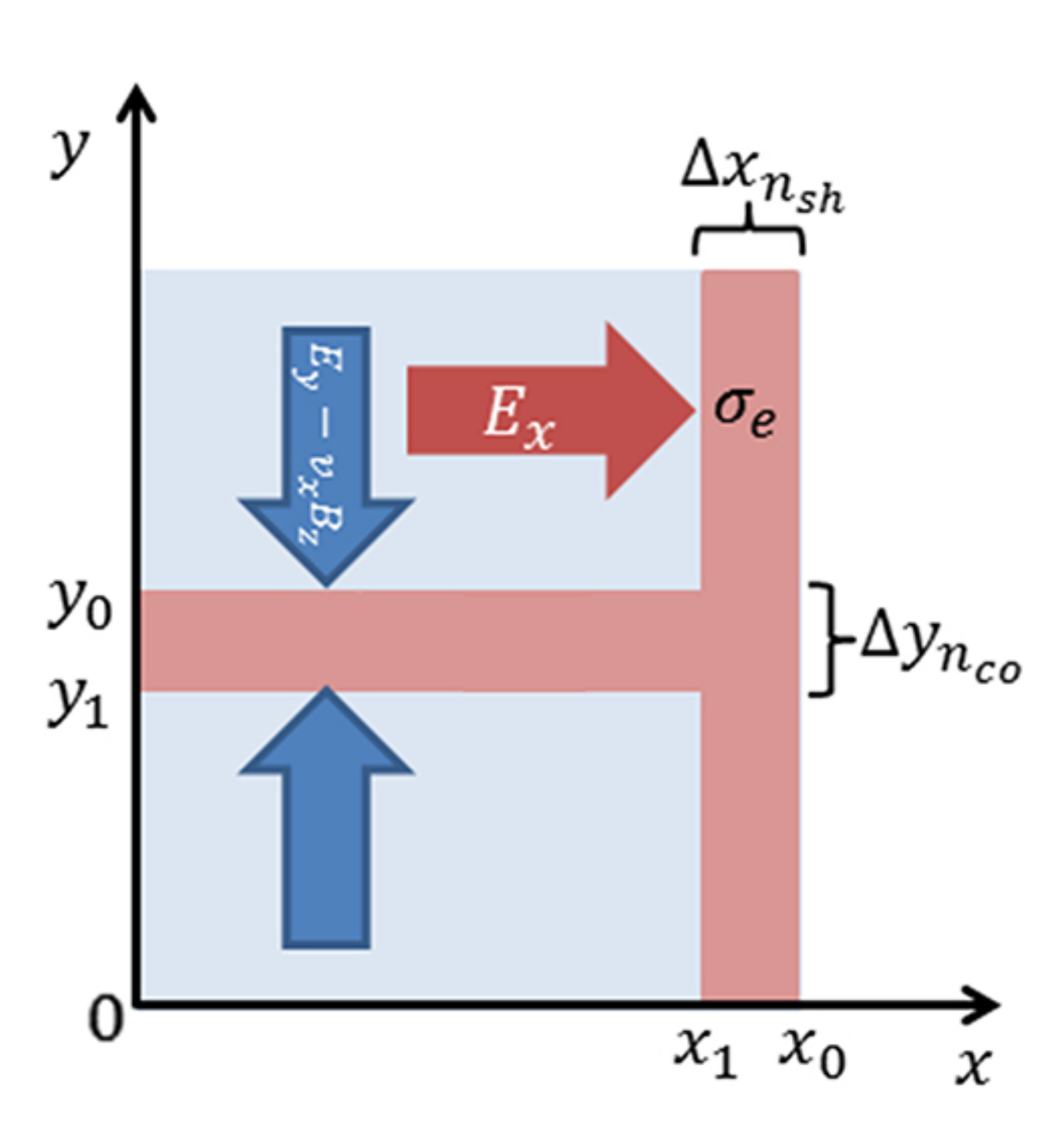}
\caption{\label{fig2} \small \textbf{A physical model of the focusing and accelerating of the $\mu^+$ beam in a donut wakefield driven by a LG laser pulse.} The physical model is in the range of the red dotted line shown in Figure 1(c). The center red rectangle represents the electron column at the center of the donut wakefield. The right red rectangle represents the electron sheath at the front of the donut wakefield. The density of the electron column and the electron sheath are assumed to be uniform. The blue arrow is the direction of the forcing force. The red arrow is the direction of the longitudinal wakefield. $\sigma_e = n_{sh}\Delta x_{n_{sh}}$, where $n_{sh}$ and $\Delta x_{n_{sh}}$ are the density and the width of the electron sheath at the front of the wakefield.}
\end{figure*}

We propose a simple mode shown in Figure \ref{fig2} to illustrate the focusing and accelerating of the $\mu^+$ beam in the nonlinear donut wakefield driven by the LG laser. The longitudinal electric field $E_x$ is equal to zero at $x=0$. For $0<x<x_1$, the longitudinal electric field $E_x$ is the acceleration field for the positive particles. In the range of $0 < x < x_1$,$y<y_1$ and $y>y_0$, the density of the plasma electrons is close to zero. The electron column at the center line of the donut wakefield provides the forcing force for the positive particles. The electron sheath at the front of the wakefiled provides the acceleration filed for positive particles. The Gauss's law can be expressed as:
\begin{equation}
\oiint_s\overrightarrow{E}d\overrightarrow{S}=\frac{q}{\varepsilon_0},   \label{eq3}
\end{equation}
where $\oiint_s$ is the integral of the closed surface S, $\overrightarrow{E}$ is the electric field, q is the charge inner the closed surface S, $\varepsilon_0$ is the permittivity of vacuum. For $0<x<x_0$, $y<y_1$ and $y>y_0$, the longitudinal acceleration field can be expressed as:
\begin{equation}
E_x=[-\int_{x_0}^x n_e  dx-n_p (x_0-x)] \frac{e}{\varepsilon_0 } +E_{x_0}, \label{eq4}
\end{equation}
where $n_e$ is the electron density, $n_p$ is the proton density considered as a constant, e is the elementary charge. $E_{x_0}$ is equal to zero. It is assumed that the density of the electron sheath at the front of the donut wakefield is a constant $n_{sh}$. The $\sigma_e$ is defined as $\sigma_e \equiv n_{sh}\Delta x_{n_{sh}}$, where $\Delta x_{n_{sh}}$ is the width of this electron sheath. For $0<x<x_1$,$y<y_1$ and $y>y_0$ close to the central part, the longitudinal acceleration field $E_x$  can be simplified as:
\begin{equation}
E_x=[\sigma_e-n_p (x_0-x)] \frac{e}{\varepsilon_0}. \label{eq5}
\end{equation}
Eq.(\ref{eq5}) shows that the acceleration field $E_x$ is proportional to $\sigma_e$.

\begin{figure*}
\includegraphics[scale=0.6]{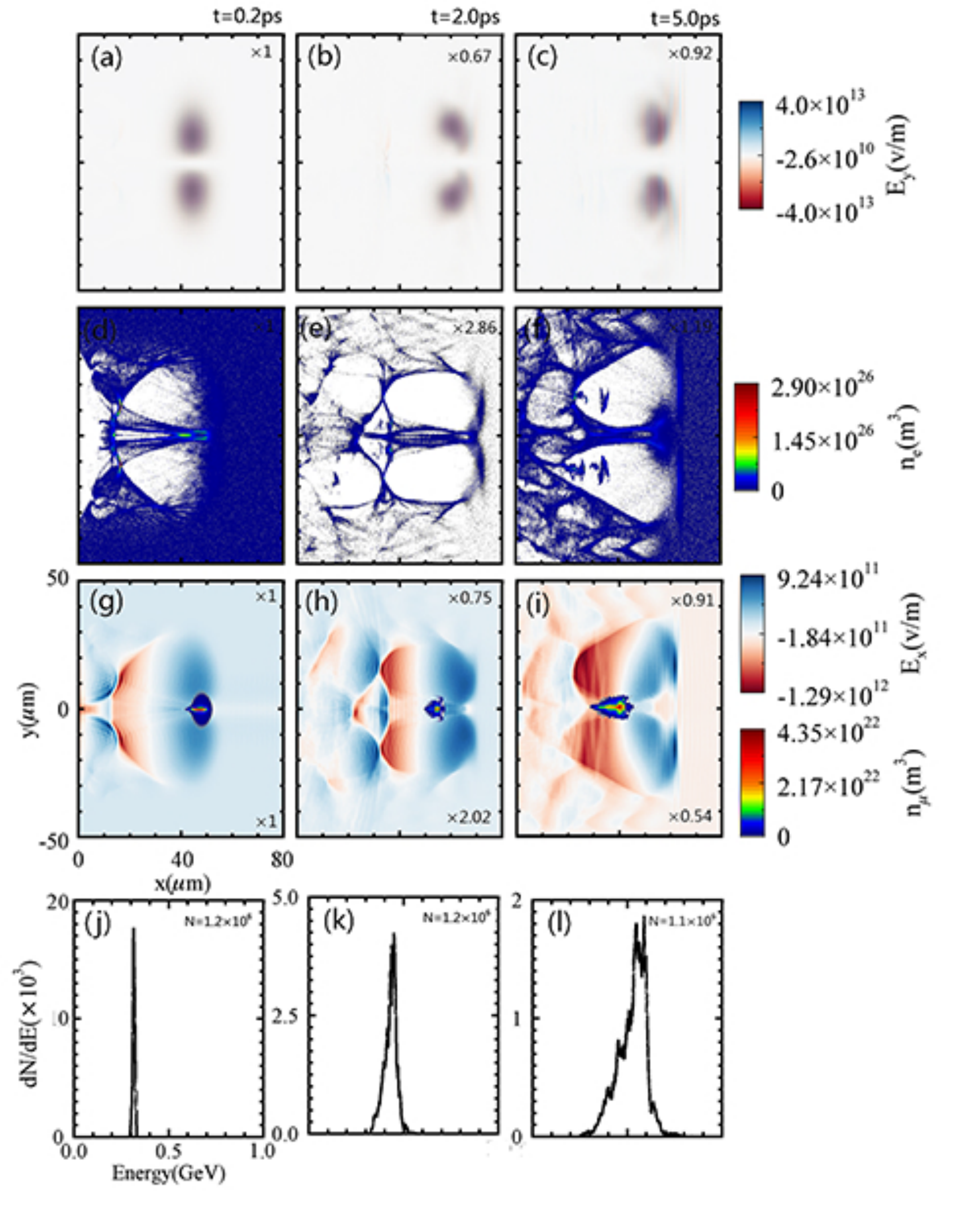}
\caption{\label{fig3} \small \textbf{The acceleration process of the $\mu^+$ bunch in a donut wakefield driven by a LG laser pulse.} This snapshots are taken at the beginning, ongoing and ending of the acceleration time $t = 0.2ps$, the $t = 2.0ps$ and $t =5.0ps$. (a-i) are the longitudinal slices of the simulation box at the plane of $z = 0\mu m$. (a-c) show the evolution of the laser pulse. (d-f) show the longitudinal structure of the donut wakefield. The (g-i) are the distributions of the longitudinal wakefield $E_x$ and the $\mu^+$ density. (k-l) are the energy spectrums of the $\mu^+$ beam at the corresponding moment. }
\end{figure*}

 The three-dimensional PIC simulation's results in Figure \ref{fig3} show the focusing and accelerating process of the $\mu^+$ beam in the donut wakefield driven by the LG laser. Figure \ref{fig3} show the simulation results using a uniform plasma with density $n_e=5\times10^{24}m^3$ and the LG laser with the $LG_{(0,1)}$ model, the spot size $\omega_0=15\mu m$ , the normal intensity $a_0=14$, and the pulse width $\tau=25\mathrm{fs}$. At the beginning of the simulation, a monoenergetic $\mu^+$ bunch with initial energy $E_0=\mathrm{300MeV}$ is placed at the front part of the acceleration filed where the laser exists shown in Figure \ref{fig3} (g). The response of the $\mu^+$ to the laser oscillating field is slower than the electron. During the acceleration process the moun beam can be located at the range of the LG laser exists, which ensures that the $\mu^+$ bunch has the longer acceleration length. The detailed simulation parameters are shown in Method.

 Figures \ref{fig3}(a-c) show that the LG laser pulse can be self-guided, which is attributed to the distribution of the refractive index at the front of the donut wakefield. Figures \ref{fig3}(d-f) show that the donut wakefield propagate in plasma stably within $\mathrm{5ps}$, providing the continuous accelerating and focusing filed for the $\mu^+$ bunch. The acceleration fields of the two spherical wakefields are separated at $\mathrm{5ps}$, which can be explained that the electron column at the center line of the donut wakefield retroact on the laser pulse. Figures \ref{fig3}(g-i) show that due to the electron column at the center line of the donut wakefield, the transversal size of the $\mu^+$ beam is focused from initial $\omega_0=5\mu m$ to $\omega=1\mu m$ within several picoseconds. The $\mu^+$ bunch with initial energy 300MeV goes back relative to the donut wakefield during the acceleration process. At $\mathrm{5ps}$ the $\mu^+$ beam is located at the decelerating filed. And the $\mu^+$ acceleration is completed. The final peak energy of the $\mu^+$ bunch in the dount wakefield is about $\mathrm{500MeV}$.

 \begin{figure*}
\includegraphics[scale=0.6]{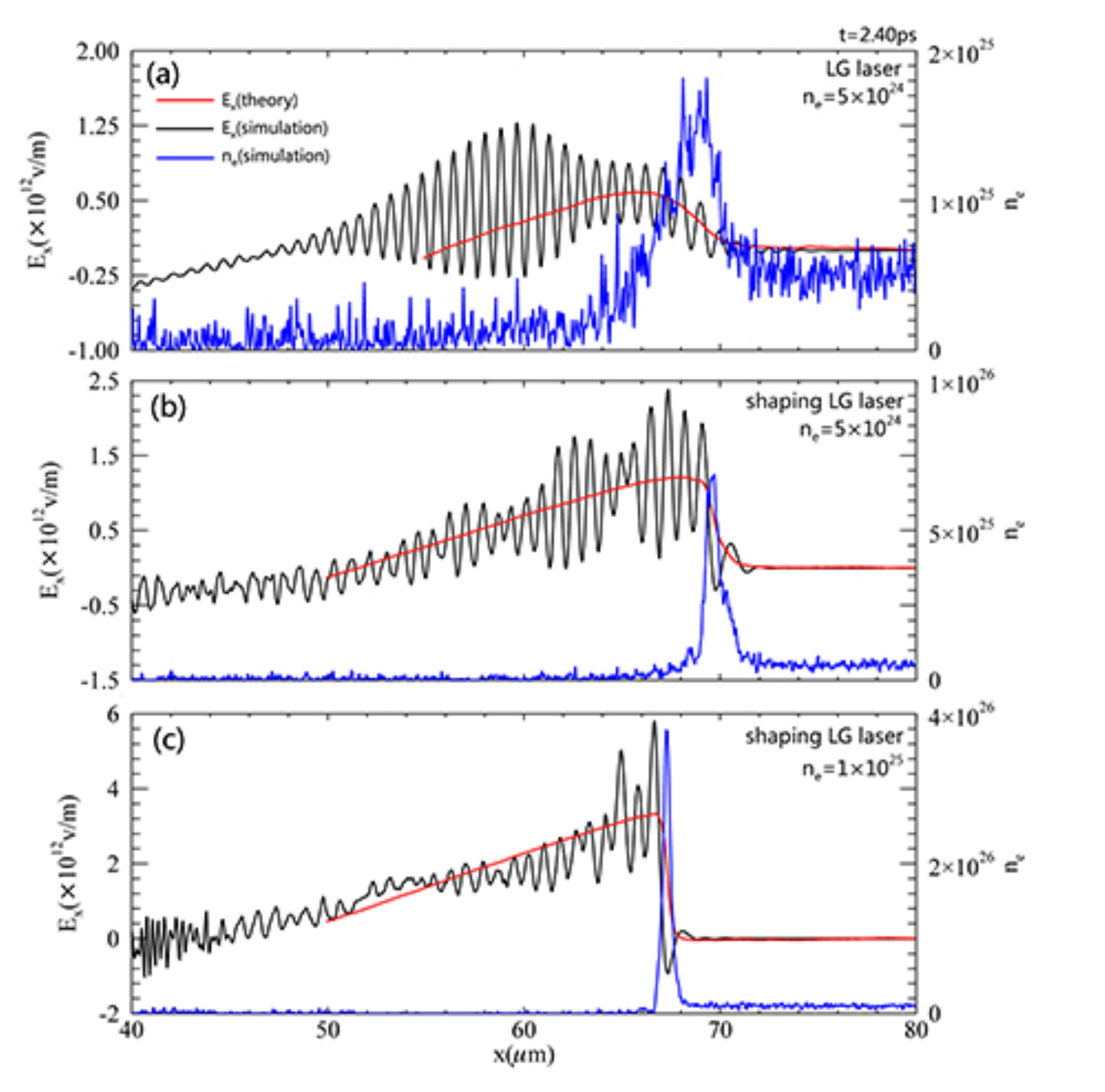}
\caption{\label{fig4} \small \textbf{The comparison of the wakefields driven by a LG laser pulse (a) and a shaped LG laser pulse (b-c) illustrating that the shaped LG laser can drive the higher acceleration filed for the positive particles.} Blue line is the simulation result of the electron density. The black  and red lines are the simulation and theory results of the acceleration filed $E_x$. Those three lines are both located at the $y=0\mu m$, $z=8\mu m$. There are three donut wakefields driven by the LG laser in a plasma with density $n_e=5\times10^{24}m^3$ (a), the shaped LG laser in a plasma with density $n_e=5\times10^{24}m^3$ (b) and the shaped LG laser in a plasma with density  $n_e=1\times10^{25}m^3$ (c).}
\end{figure*}

A shaped LG laser pulse can push the plasma electrons form an electron sheath with larger $\sigma_e$ at the front of the donut wakefield and enlarges the acceleration field. the shaping of the general gauss laser is proposed by H.W.Wang et.al.\textsuperscript{\cite{wang2011laser37}} They demonstrate that as the relativistic self-focusing(RSF)\textsuperscript{\cite{chen1993necessary38}}\textsuperscript{\cite{pukhov1996relativistic39}}, the relativistic self-phase modulation (RSPM)\textsuperscript{\cite{max1974self40}}\textsuperscript{\cite{shorokhov2003self41}} and the relativistic transparency occur in the laser interaction with the near-critical plasma, there are three shaping effects: the laser intensity enhancement, the laser profile steepening, and the absorption of nonrelativistic prepulse. Our simulation's results show that the LG laser pulse also can be shaped by the near-critical plasma. The degree of the shaping is controlled by the length of the near-critical plasma. In our simulations, the length of the near-critical plasma is $5\mu m$. The blue lines in Figures \ref{fig4} are the one-dimensional density distributions of the plasma electrons located at $y=0\mu m$, $z=4\mu m$ in the three-dimensional simulation box. The simulation results in Figures \ref{fig4}(a-c) use the same LG laser pulses and the different plasma conditions. The shaping of the LG laser is completed by a near-critical-density plasma. The blue line in Figure \ref{fig4}(a) shows the simulation result in a uniform plasma with the density $n_e=5\times10^{24}m^3$. The blue line in Figure \ref{fig4}(b) shows the simulation result using two plasma layers. The first layer is the near-critical-density plasma of $5\mu m$ length with the density $n_e=2.67\times10^{27}m^3$ , located at $3\mu m<x<8\mu m$ and used to shape the LG laser pulse. The second layer with density $n_e=5\times10^{24}m^3$ is filled the range of $x<40\mu m$. The blue line in Figure \ref{fig4}(c) shows the simulation result also using two plasma layers. The first layer is same as above list. The second plasma layer with density $n_e=1\times 10^{25}m^3$ is also filled the range of $x<40\mu m$. Compared with the blue line in Figure \ref{fig4}(a), Figure \ref{fig4} (b) shows that the electron sheath generated by the shaped laser pulse has the larger $\sigma_e$. Further more, the blue line in Figure \ref{fig4}(c) shows that in the plasma with the higher density, the $\sigma_e$ of the electron sheath is larger. The black lines in the Figures \ref{fig4} show the simulation results of the one-dimensional acceleration filed $E_x$ located at $y=0\mu m$, $z=4\mu m$ in the three-dimensional simulation box. Corresponding to the black lines, the red lines in the Figures \ref{fig4} show the theory results of the Eq. (\ref{eq4}). The electron density $n_e$ of the Eq. (\ref{eq4}) can be obtained by the blue lines of Figure \ref{fig4}. The theory results agree with the simulation results. Compared with the red line in Figure \ref{fig4}(a), Figure \ref{fig4} (b) shows that the shaped laser pulse can drive the higher acceleration filed. Further more, the red line in Figure \ref{fig4}(c) shows that in the plasma with the higher density, the acceleration field is larger and up to about $3.5\mathrm{TV/m}$.

\begin{figure*}
\includegraphics[scale=0.6]{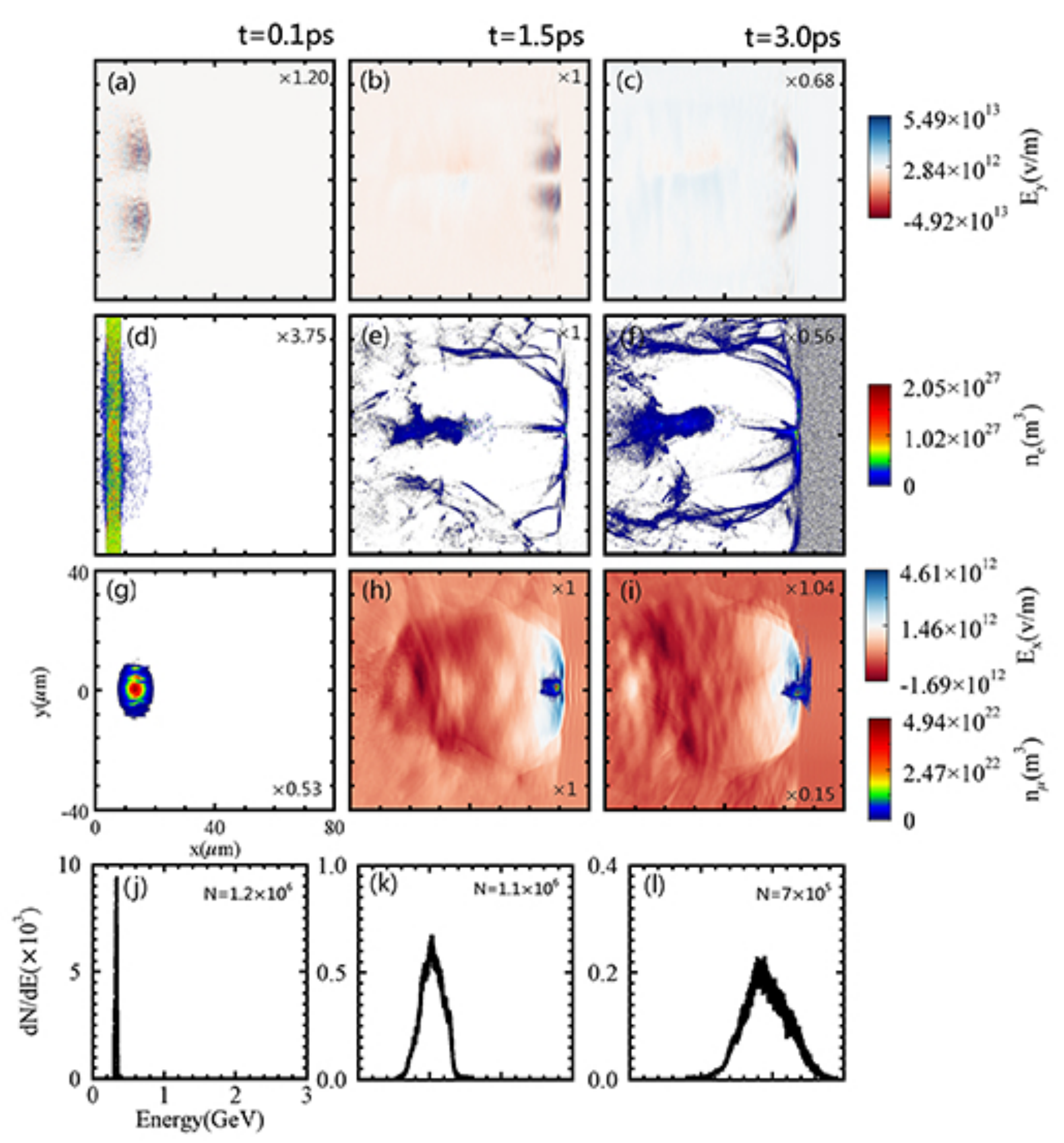}
\caption{\label{fig5} \small \textbf{The acceleration process of the $\mu^+$ bunch in a donut wakefield driven by a shaped LG laser pulse.} This snapshots are taken at the beginning, ongoing and ending of the acceleration time $t = 0.1ps$,  $t = 1.5ps$ and $t =3.0ps$. (a-i) are the longitudinal slices of the simulation box at the plane of $z = 0\mu m$.  (a) shows the shaping of the LG Laser in the near-critical-density plasma. (b), (c) show the evolution of the laser pulse in a plasma with the density $n_e=1\times10^{25}m^3$. (d) is the near-critical-density plasma. (e), (f) show the longitudinal structure of the donut wakefield. (g) is the initial density distribution of $\mu^+$ bunch. (h), (i) are the distributions of the longitudinal field $E_x$ and the $\mu^+$ density. (j-l) are the energy spectrums of the $\mu^+$ beam at the corresponding moment.}
\end{figure*}

 Figure \ref{fig5} show the evolution of the $\mu^+$ acceleration process in the donut wakefiled driven by the shaped LG laser in detail. The laser parameters are same as those of Figure \ref{fig3}. There are two plasma layers in this simulation. The first layer is the near-critical-density plasma of $5\mu m$ length with the density $n_e=2.67\times10^{27} m^3$, located at $3\mu m<x<8\mu m$ and used to shape the LG laser pulse, the second layer with the density $n_e=1\times10^{25}m^3$ is filled the range of $x<40\mu m$. Adjusting the injection time of the $\mu^+$ beam ensures that at the beginning of the simulation the $\mu^+$ bunch is located at the front part of the acceleration field. The detail simulation parameters are list in method.

 Figure \ref{fig5} show that in a plasma with density $n_e=1\times10^{25}m^3$, the shaped LG laser pulse also can drive the acceleration filed at the center line of the donut wakefiled for the positive particles. Although the electron column at the center line of the donut wakefiled is unstable during the acceleration process, the transversal size of the $\mu^+$ beam is focused. Figure \ref{fig5}(a) shows that the LG laser pulse is shaped by the near-critical-density plasma. The shaped LG laser has the shorter rise time $r_t=13fs$ and the higher amplitude $E_y=5.6\times10^{13} V/m$, compared the LG laser shown in Figure \ref{fig3}(a) with the rise time $r_t=33fs$ and the amplitude $E_y=4\times10^{13} V/m$. Figure \ref{fig5}(b), (c) show that the LG laser pulse is self-guided. A large proportion of the laser's energy is consumed at $t=3ps$. Figure \ref{fig5}(e), (f) show that the donut wakefield propagate in the plasma stablely within $\mathrm{3ps}$, providing the continuous accelerating and focusing fileds for the $\mu^+$ bunch. The acceleration gradient is up to $4.5\times10^{12} V/m$ at $t=1\mathrm{ps}$. Figures \ref{fig5}(g-i) shows that due to the electron column at the center line of the donut wakefiled, the transversal size of the $\mu^+$ beam is focused from initial $\omega_0=5\mu m$ to $\omega=1\mu m$ within several picoseconds. At the beginning of the acceleration, the $\mu^+$ bunch with initial energy $300\mathrm{MeV}$ goes back relative to the donut wakefiled and be accelerated. When the velocity of the $\mu^+$ bunch is larger than that of the donut wakefiled, the $\mu^+$ bunch goes forward relative to the donut wakefiled. Finally the $\mu^+$ bunch exceedes the dount wakefiled and the acceleration process is completed. The final energy gain of the $\mu^+$ bunch at the donut wakefiled driven by the shaped LG laser pulse is about $\mathrm{1.5GeV}$.

 \begin{figure*}
\includegraphics[scale=0.6]{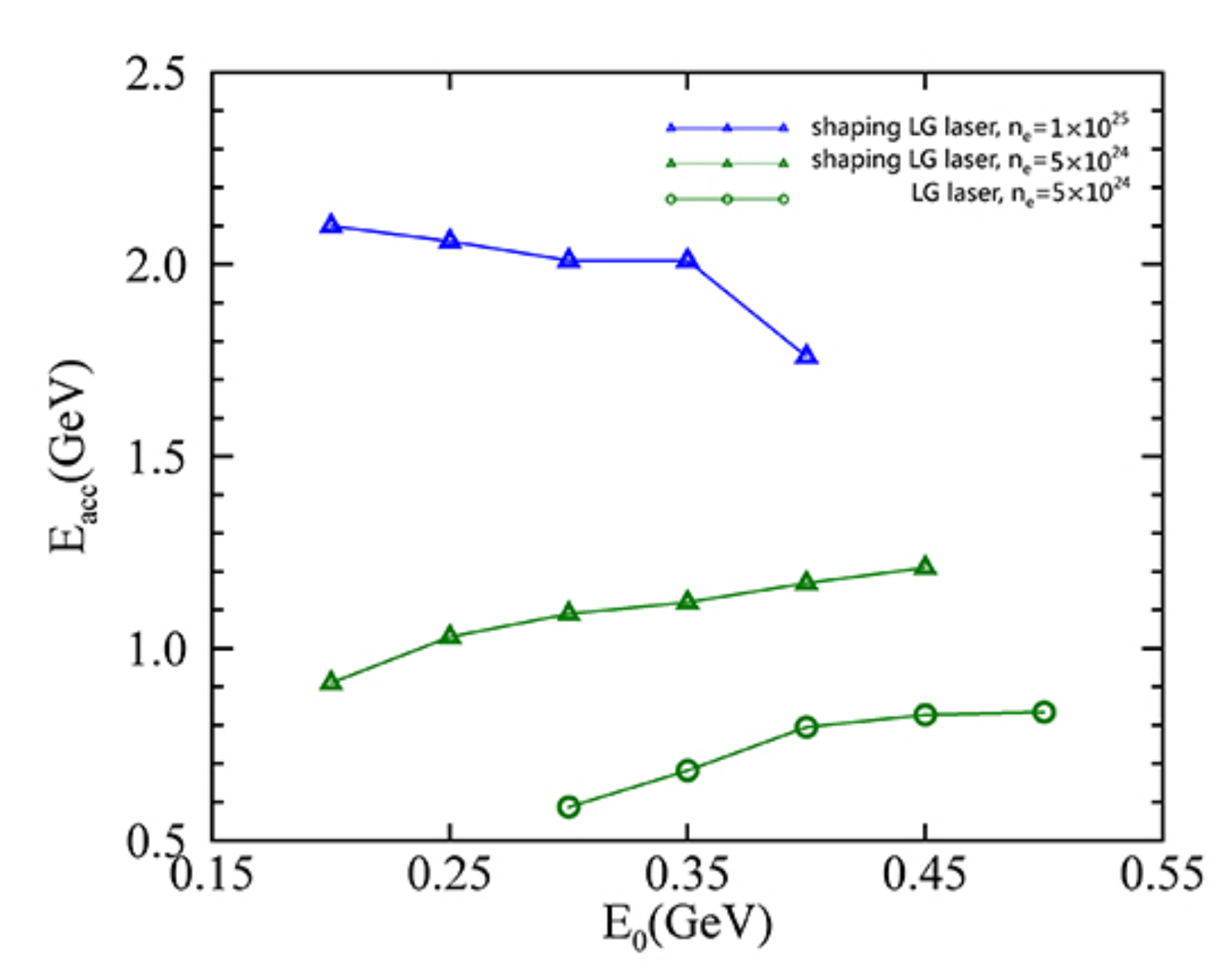}
\caption{\label{fig6} \small \textbf{The relationship between the initial energy $E_0$ and the final acceleration energy $E_{acc}$ of the $\mu^+$ beam in the donut wakefileds with the three different plasma conditions and the same LG laser pules.} The shaping of the LG laser is completed by a near-critical-density plasma. We choose the maximum peak energy during the acceleration process as the $E_{acc}$. The blue triangle, green triangle, green circle are the simulation results using the shaped LG laser and $n_e=1\times10^{25}$, the shaped LG laser and $n_e=5\times10^{24}$, the LG laser and $n_e=5\times10^{24}$.}
\end{figure*}

 Figure \ref{fig6} shows the relationship between the initial energy $E_0$ and the final acceleration energy $E_{acc}$ of the $\mu^+$ beam in the donut wakefileds with the three different plasma conditions. In the donut wakefield driven by the shaped LG laser injecting to the plasma with density $n_e=1\times10^{25}m^3$ , the energy gain of the $\mu^+$ beam is increased six times than those  of the $\mu^+$ beam in the donut wakefield driven by the normal LG laser. In the donut wakefield driven by the shaped LG laser injecting the plasma with the density $n_e=1\times10^{25}m^3$, the end of the acceleration is that the $\mu^+$ bunch exceeds the wakefiled. When the initial energy of $\mu^+$ beam is higher, the acceleration length of the $\mu^+$ beam will be longer. Therefor, the energy gain of the $\mu^+$ bunch will be increased,
 accordingly. In the donut wakefield driven by the normal LG laser pulse or the shaped LG laser injecting the plasma with the density $n_e=5\times10^{25}m^3$, the end of the acceleration is that the $\mu^+$ bunch goes backward to the declaration filed. With higher initial energy, the $\mu^+$ bunch has longer acceleration length. Therefor, the energy gain of the $\mu^+$ bunch will be increased.

 In conclusion, we propose a sample physical model to illustrate that the $\mu^+$ bunch can be focused and accelerated in the donut wakefield driven by a LG laser, and calculate that the acceleration field $E_x$ is proportional to the $\sigma_e$. Figure 3  show the focusing and acceleration process of the $\mu^+$ beam in the donut wakefield driven by the LG laser. The simulation results show that the $\mu^+$ bunch energy gain is about $200\mathrm{MeV}$ and the transversal size of the $\mu^+$ beam is focused from initial $\omega_0=5\mu m$ to $\omega=1\mu m$ within several picoseconds. We propose and demonstrate that the acceleration filed of the $\mu^+$ beam can be optimized by the shaped LG laser. the Shaped LG laser can push the plasma electrons to generate the electron sheath with larger $\sigma_e$ at the front of the wakefield and enlarges the acceleration filed. This higher acceleration field can accelerate the $\mu^+$ beam from lower initial energy $E_0$ to higher acceleration energy $E_{acc}$. The $\mu^+$ beam is accelerated from $200\mathrm{MeV}$ to $2\mathrm{GeV}$ in the donut wakefield driven by the shaped LG laser pulse. Although the electron column at the center line of the donut wakefiled is unstable during the acceleration process, the transversal size of the $\mu^+$ beam is focused from initial $\omega_0=5\mu m$ to $\omega=1\mu m$ within several picoseconds.

 {\textbf{\large {\\Method}}}
   \textbf{\\The simulation parameters}. Our simulations are divided into three types: A. A normal LG laser injects the plasma with density
$n_e=5\times10^{24}m^3$. B. A Shaped LG laser injects the plasma with density $n_e=5\times10^{24}m^3$ C. A Shaped LG laser injects the plasma with density $n_e=1\times10^{25}m^3$. Table \ref{tab:table1} is the detail simulation parameters of this three types. the $gauss(x,x_0,\omega)$ function shown in Table \ref{tab:table1} can be expressed as $gauss(x,x_0,\omega )=exp-(\frac{x-x_0}{\omega})^2$

\begin{table}[b]
\caption{\label{tab:table1}
The simulation parameters in three types, We only list the parameters of type B and C different with those of type A }
\begin{ruledtabular}
\begin{tabular}{cccc}
  & A & B & C \\
 \hline
 \textbf{window parameters} & & & \\
 size & $x\times y\times z=80\mu m\times 100\mu m\times 100\mu m$ & $x=80\mu m$ & $x=80\mu m$ \\
 The number of grids &  $x\times y\times z=1600\times 200\times 200$ & & \\
 velocity & $v=2.9999 \times 10^8 m/s$ & & \\
 moving start time & $t=270fs$ & $t=295fs$ & $t=270fs$ \\
 \hline
 \textbf{Plasma parameters} & & & \\
 Plasma particles & electron and proton & & \\
 density & $n_e=5\times10^{24}m^3$ & Two plasma layer & Two plasma layer \\
 & & a.$n_e=2.67\times10^{27}m^3$ & a.$n_e=2.67\times10^{27}m^3$\\
 & & b.$n_e=5\times10^{24}m^3$ & b.$n_e=1\times10^{25}m^3$ \\
 position & Fill the simulation box & a.$8\mu m>x>3\mu m$ & a.$8\mu m>x>3\mu m$ \\
 & & b.$x>40\mu m$ &  b.$x>40\mu m$ \\
 \hline
 \textbf{Laser parameters} & & & \\
 Wave length & $0.8\mu m$ & & \\
 Intensity & $a_0=14$ & & \\
 Spot size & $\omega_0=15\mu m$ & & \\
 Profile at time axis & $t_{profile} = gauss(time, 50fs, 25fs)$ & & \\
 \hline
 \textbf{Muon beam's parameters} & & & \\
 maximum density & $n_0=3\times10^{21}$ & & \\
 Density distrubution & $n_\mu=n_0*gauss(time,35fs,10fs)$ & $n_\mu=n_0*gauss(time,50fs,10fs)$ & $n_\mu=n_0*gauss(time,50fs,10fs)$ \\
 & & $*gauss(y,0,5\mu m)$ & $*gauss(y,0,5\mu m)$ \\
  & & $*gauss(z,0,5\mu m)$ & $*gauss(z,0,5\mu m)$ \\
  Total number & $N=1.2\times 10^6$ & & \\
  temperature & 0eV & & \\

\end{tabular}
\end{ruledtabular}
\end{table}

{\textbf{\large {\\Data availability}}}. The data that support the findings of this study are available
from the corresponding authors on request.

\begin{acknowledgments}
This work was supported in part by Innovation Project of IHEP (542017IHEPZZBS11820, 542018IHEPZZBS12427); the CAS Center for Excellence in Particle Physics (CCEPP); the Meritocracy Research Funds of China West Normal University(No. 17YC504);the National Key $R\&D$ Program of China (Grant No. 2018YFA0404802), National Natural Science Foundation of China (Grant No. 11875319), the Hunan Provincial Science and Technology Program (Grant No. 2020RC4020).
\end{acknowledgments}


\nocite{*}

\begin{thebibliography}{99}


\bibitem{berger2014mu3e1} N. Berger, M. Collaboration, et al., The mu3e experiment, Nuclear Physics B-Proceedings Supplements \textbf{248}, 35(2014).
\bibitem{bartoszek2015mu2e2} L. Bartoszek, E. Barnes, J. Miller, J. Mott, A. Palladino, J. Quirk, B. Roberts, J. Crnkovic, V. Polychronakos, V. Tishchenko, et al., Mu2e technical design report,  arXiv preprint arXiv:1501.05241 (2015).
\bibitem{kutschke2009mu2e3} R. K. Kutschke, The mu2e experiment at fermilab, in AIP Conference Proceedings, Vol. 1182 (2009) pp. 718-721.
\bibitem{grassi2005meg4} M. Grassi, M. Collaboration, et al., The meg experiment at psi: status and prospect, Nuclear Physics B-Proceedings Supplements \textbf{149}, 369 (2005).
\bibitem{kuno2013search5} Y. Kuno, A search for muon-to-electron conversion at j-parc: the comet experiment, Progress of Theoretical and Experimental Physics 2013, 022C01 (2013).
\bibitem{farley20044756} F. Farley and Y. K. Semertzidis, The 47 years of muon g-2, Nuclear Physics B-Proceedings Supplements 52, 1 (2004).
\bibitem{charpak1961measurement7} G. Charpak, F. Farley, R. Garwin, T. Muller, J. Sens, V. Telegdi, and A. Zichichi, Measurement of the anomalous magnetic moment of the muon, Nuclear Physics B-Proceedings Supplements \textbf{6}, 128 (1961).
\bibitem{bailey1979final8} J. Bailey, K. Borer, F. Combley, H. Drumm, C. Eck, F. Farley, J. Field, W. Flegel, P. Hattersley, F. Krienen, et al., Measurement of the anomalous magnetic moment of the muon, Nuclear Physics B \textbf{150}, 1 (1979). 
\bibitem{neutrinofac9} G. Charpak, F. Farley, R. Garwin, T. Muller, J. Sens, V. Telegdi, and A. Zichichi, Interim design report, arXiv preprint arXiv:1112.2853 (2011).
\bibitem{cao2014muon10} J. Cao, M. He, Z.-L. Hou, H.-T. Jing, Y.-F. Li, Z.-H. Li, Y.-P. Song, J.-Y. Tang, Y.-F. Wang, Q.-F. Wu, et al., Muon-decay medium-baseline neutrino beam facility, Physical Review Special Topics-Accelerators and Beams \textbf{17}, 090101 (2014).
\bibitem{ankenbrandt1999status11}C. M. Ankenbrandt, M. Atac, B. Autin, V. I. Balbekov, V. D. Barger, O. Benary, J. S. Berg, M. S. Berger, E. L. Black, A. Blondel, $et$ $al.$, Status of muon collider research and development and future plans, Physical Review Special Topics-Accelerators and Beams \textbf{2}, 081001 (1999).
\bibitem{barger1995s} V. Barger, M. Berger, J. Gunion, and T. Han, s-channel higgs boson production at a muon-muon collider, Physical Review Letters \textbf{75}, 1462 (1995).
\bibitem{wang2001nuclear12} Y. Wang, Z. Pan, Y. Ho, Y. Xu, and A. Du, Nuclear instruments and methods in physics research section b: beam interactions with materials and atoms, Nuclear Instruments and Methods in Physics Research B \textbf{180}, 251 (2001). 
\bibitem{sonier2000musr13} J. E. Sonier, J. H. Brewer, and R. F. Kiefl, $\mu sr$ studies of the vortex state in type-ii superconductors, Reviews of Modern Physics \textbf{72}, 769 (2000). 
\bibitem{amato1997heavy14} A. Amato, Heavy-fermion systems studied by $\mu sr$ technique, Reviews of Modern Physics \textbf{69}, 1119 (1997). 
\bibitem{roduner2012positive15} E. Roduner, $The positive muon as a probe in free radical chemistry: potential and limitations of the \mu SR techniques$, Vol. 49 (Springer Science and Business Media, 2012). 
\bibitem{olbert1954production20} S. Olbert, Production spectra of cosmic-ray mesons in the atmosphere, Physical Review \textbf{96}, 1400 (1954). 
\bibitem{bose1944cosmic21} D. Bose, B. Choudhuri, and M. Sinha, Cosmic-ray meson spectra, Physical Review \textbf{65}, 341 (1944).
\bibitem{miyake2009j22} Y. Miyake, K. Nishiyama, N. Kawamura, P. Strasser, S. Makimura, A. Koda, K. Shimomura, H. Fujimori, K. Nakahara, R. Kadono, $et$ $al.$, J-parc muon source, muse, Nuclear Instruments and Methods in Physics Research Section A: Accelerators, Spectrometers, Detectors and Associated Equipment \textbf{600}, 22 (2009).
\bibitem{morenzoni2000low23} E. Morenzoni, H. Gluckler, T. Prokscha, H. Weber, E. Forgan, T. Jackson, H. Luetkens, C. Niedermayer, M. Pleines, M. Birke, $et$ $al.$, Low-energy musr at psi: present and future, Physica B: Condensed Matter \textbf{289}, 653 (2000).
\bibitem{matsuzaki2001riken24} T. Matsuzaki, K. Ishida, K. Nagamine, I. Watanabe, G. Eaton, and W. Williams, The riken-ral pulsed muon facility, Nuclear Instruments and Methods in Physics Research Section A: Accelerators, Spectrometers, Detectors and Associated Equipment \textbf{465}, 365 (2001).
\bibitem{leemans2014multi25} W. Leemans, A. Gonsalves, H.-S. Mao, K. Nakamura, C. Benedetti, C. Schroeder, C. Tth, J. Daniels, D. Mittelberger, S. Bulanov, $et$ $al.$, Multi-gev electron beams from capillary-discharge-guided subpetawatt laser pulses in the self-trapping regime, Physical review letters \textbf{113}, 245002 (2014).
\bibitem{chen1985acceleration26} P. Chen, J. Dawson, R. W. Huff, and T. Katsouleas, Acceleration of electrons by the interaction of a bunched electron beam with a plasma, Physical review letters \textbf{54}, 693 (1985).
\bibitem{litos201427} M. Litos, E. Adli, W. An, C. Clarke, C. Clayton, S. Corde, J. Delahaye, R. England, A. Fisher, J. Frederico, $et$ $al.$, High-efficiency acceleration of an electron beam in a plasma wakefield accelerator, Nature \textbf{515}, 92 (2014).
\bibitem{hogan2010plasma28} M. Hogan, T. Raubenheimer, A. Seryi, P. Muggli, T. Katsouleas, C. Huang, W. Lu, W. An, K. Marsh, W. Mori, $et$ $al.$, Plasma wakefield acceleration experiments at facet, New Journal of Physics \textbf{12}, 055030 (2010).
\bibitem{zhang2018all29} F. Zhang, Z. Deng, L. Shan, Z. Zhang, B. Bi, D. Liu, W. Wang, Z. Yuan, C. Tian, S. Yang, $et$ $al.$, All-optical acceleration in the laser wakefield, High Power Laser Science and Engineering \textbf{6}, 693 (2018).
\bibitem{vieira2014nonlinear30} J. Vieira and J. Mendonca, Nonlinear laser driven donut wakefields for positron and electron acceleration, Physical Review Letters \textbf{112}, 215001 (2014).
\bibitem{mendoncca2014donut31} J. Mendonca and J. Vieira, Donut wakefields generated by intense laser pulses with orbital angular momentum, Physics of Plasmas \textbf{21}, 033107 (2014). 
\bibitem{zhang2016acceleration32} G.-B. Zhang, M. Chen, C. Schroeder, J. Luo, M. Zeng, F.-Y. Li, L.-L. Yu, S.-M. Weng, Y.-Y. Ma, T.-P. Yu, et al., Acceleration and evolution of a hollow electron beam in wakefields driven by a laguerre-gaussian laser pulse, Physics of Plasmas \textbf{23}, 033114 (2016).
\bibitem{allen1992orbital33} L. Allen, M. W. Beijersbergen, R. Spreeuw, and J. Woerdman, Orbital angular momentum of light and the transformation of laguerre-gaussian laser modes, Physical review A \textbf{45}, 8185 (1992).
\bibitem{arber2015contemporary34} T. Arber, K. Bennett, C. Brady, A. Lawrence-Douglas, M. Ramsay, N. Sircombe, P. Gillies, R. Evans, H. Schmitz, A. Bell, $et$ $al.$, Contemporary particle-in-cell approach to laser-plasma modelling, Plasma Physics and Controlled Fusion \textbf{57}, 113001 (2015).
\bibitem{lu2006nonlinear35} W. Lu, C. Huang, M. Zhou, M. Tzoufras, F. Tsung, W. Mori, and T. Katsouleas, A nonlinear theory for multidimensional relativistic plasma wave wakefields, Physics of Plasmas \textbf{13}, 056709 (2006).
\bibitem{lu2007generating36} W. Lu, M. Tzoufras, C. Joshi, F. Tsung, W. Mori, J. Vieira, R. Fonseca, and L. Silva, Generating multi-gev electron bunches using single stage laser wakefield acceleration in a 3d nonlinear regime, Physical Review Special Topics-Accelerators and Beams \textbf{10}, 061301 (2007). 
\bibitem{wang2011laser37} H. Wang, C. Lin, Z. Sheng, B. Liu, S. Zhao, Z. Guo, Y. Lu, X. He, J. Chen, and X. Yan, Laser shaping of a relativistic intense, short gaussian pulse by a plasma lens, Physical review letters \textbf{107}, 265002 (2011). 
\bibitem{chen1993necessary38}X. Chen and R. Sudan, Necessary and sufficient conditions for self-focusing of short ultraintense laser pulse in underdense plasma, Physical review letters \textbf{70}, 2082 (1993). 
\bibitem{pukhov1996relativistic39} A. Pukhov and J. Meyer-ter Vehn, Relativistic magnetic self-channeling of light in near-critical plasma: three-dimensional particle-in-cell simulation, Physical review letters \textbf{76}, 3975 (1996).
\bibitem{max1974self40} C. E. Max, J. Arons, and A. B. Langdon, Self-modulation and self-focusing of electromagnetic waves in plasmas, Physical review letters \textbf{33}, 209 (1974). 
\bibitem{shorokhov2003self41} O. Shorokhov, A. Pukhov, and I. Kostyukov, Self-compression of laser pulses in plasma, Physical review letters 91, 265002 (2003). 
\end{thebibliography}
\end{document}